\documentclass{ws-ijqi}

\def\draftnote{}
\def\trimmarks{}

\long\def\dummy#1{}
\let\ds\displaystyle

\usepackage{amstext,boldmath}

\let\ds\displaystyle
\long\def\magma#1{}
\long\def\dummy#1{}
\def\email#1{{\normalfont\texttt{#1}}}

\def\wgt{\mathop{\rm wgt}}
\def\tr{\mathop{\rm tr}\nolimits}

\def\N{{\mathbb N}}

\def\BibTeX{{\rm B\kern-.05em{\sc i\kern-.025em b}\kern-.08em
    T\kern-.1667em\lower.7ex\hbox{E}\kern-.125emX}}

\def\QECC(#1,#2,#3,#4){[\![#1,#2,#3]\!]_{#4}}

\begin{document}

\def\mytitle{On optimal quantum codes}

\def\myabstract{
We present families of quantum error-correcting codes which are
optimal in the sense that the minimum distance is maximal.  These
maximum distance separable (MDS) codes are defined over
$q$-dimensional quantum systems, where $q$ is an arbitrary prime
power. It is shown that codes with parameters $\QECC(n,n-2d+2,d,q)$
exist for all $3\le n\le q$ and $1\le d\le n/2+1$.  We also present
quantum MDS codes with parameters $\QECC(q^2,q^2-2d+2,d,q)$ for $1\le
d\le q$ which additionally give rise to shortened codes
$\QECC(q^2-s,q^2-2d+2-s,d,q)$ for some $s$.  
}

\def\mykeywords{Quantum error-correcting codes, quantum MDS codes}

\markboth{M. Grassl, Th. Beth, M. R\"otteler}{\mytitle}

%
\catchline{}{}{}{}{}
%

\title{\mytitle}

\author{Markus Grassl, Thomas Beth}

\address{
Institut f{\"u}r Algorithmen und Kognitive Systeme,\\
Universit{\"a}t Karlsruhe, Am Fasanengarten 5, 76\,128 Karlsruhe,
Germany\\
\email{grassl@ira.uka.de}, \email{EISS\_Office@ira.uka.de}}

\author{Martin R\"otteler}

\address{
Department of Combinatorics and Optimization, Faculty of Mathematics,\\
University of Waterloo, Waterloo, Ontario, Canada, N2L 3G1\\
\email{mroetteler@iqc.ca}
}
\footnotetext{Accepted for publication in the International Journal of
Quantum Information.\\
This work was presented in part at the ERATO Conference on Quantum
Information Science (Kyoto, Japan, 2003).
}

\maketitle


\begin{abstract}
\myabstract
\end{abstract}


\section{Introduction}
In this paper, we consider error-correcting codes for quantum systems
which are composed of subsystems of dimension $p^m$, where $p$ is
prime and $m\in\N$. As a shorthand, we will use the term ``qudit''. In
the theory of classical error-correcting codes it is well known that
by increasing the size of the underlying alphabet, codes with better
parameters can be constructed.\cite{MS77,TVZ82} We will show that the
same is true for quantum error-correcting codes.

Quantum codes for qudit systems have been studied
before\cite{AhBO97,Got98:NASA,AsKn01,Ham02} including efficient
algorithms for encoding these codes.\cite{GRB03} It is known that
codes encoding one qudit into five qudits which are capable to correct
one error, denoted by $\QECC(5,1,3,q)$, exist for quantum systems of
any dimension.\cite{Chau97b} In general, by $\QECC(n,k,d,q)$ we will
denote a quantum error-correcting code (QECC) which encodes $k$ qudits
of a $q$-dimensional quantum system into $n$ qudits.  The parameter
$d$ is the minimum distance of the code.  A QECC with minimum distance
$d$ can be used to detect errors that involve at most $d-1$ of the $n$
subsystems.  Alternatively, one can correct errors that involve less
than $d/2$ subsystems.

Recently it was shown that optimal quantum codes with parameters
$\QECC(6,2,3,p)$ and $\QECC(7,3,3,p)$ exist for all primes $p\ge 3$
(see Ref.~\refcite{Feng02}). There also exist quantum codes
$\QECC(p,1,(p+1)/2,p)$ encoding one qudit into many qudits which are
capable to correct more than one error.\cite{AhBO97} We show that many
more optimal quantum codes exist.  Note that in this paper we consider
only codes of finite length, and not the asymptotic performance of
codes when the length tends to infinity (for this, see, e.\,g.,
Ref.~\refcite{Ham02}).

First we recall basic constructions of QECCs from classical
codes.\cite{AsKn01,CaSh96,Ste96:error} Then we present families of
optimal classical codes suitable for these constructions.  In
Section~\ref{sec:RainsPuncturing} we address the problem of shortening
quantum codes and conclude with a table of results.

\subsection{Quantum Codes}
For completeness, we recall some constructions of quantum
error-correcting codes from classical ones.

First, on the space $(GF(q)\times GF(q))^n\equiv GF(q)^n\times
GF(q)^n$ we consider the symplectic inner product defined by
\begin{equation}\label{eq:symplectic}
(\bm{v},\bm{w})*(\bm{v}',\bm{w}'):=\bm{v}\cdot\bm{w}'-\bm{v}'\cdot\bm{w}=\sum_{i=1}^n v_i w_i'-v_i'w_i.
\end{equation}
For codes over $GF(q)\times GF(q)$ which are $GF(q)$-linear with $q^k$
codewords, denoted by $C=(n,q^k,d)_q$, we use the notation $C^*$ for
the dual code with respect to (\ref{eq:symplectic}), i.\,e.,
$$
C^*:=\{(\bm{v},\bm{w})\in GF(q)^n\times GF(q)^n \mid
\forall \bm{c}\in C\colon (\bm{v},\bm{w})*\bm{c}=0\}.
$$
A code which is contained in its dual is called self-orthogonal. These
codes can be used to construct QECCs for qudits:\cite{AsKn01}
\begin{theorem}\label{theorem:additiveQECC}
Let $C=(n,q^k)_q$ be a self-orthogonal code over $GF(q)\times GF(q)$
with $q^k$ codewords and let $d=\min\{\wgt(\bm{v})\colon\bm{v}\in
C^*\setminus C\}$. Then there exists a QECC encoding $n-k$ qudits into
$n$ qudits with minimum distance $d$, denoted by ${\cal
C}=\QECC(n,n-k,d,q)$.
\end{theorem}
For $GF(q^2)$-linear codes over $GF(q^2)$ one can also consider
duality with respect to the Hermitian inner product on $GF(q^2)^n$,
defined by
\begin{equation}\label{eq:Hermitian}
\bm{v}*\bm{w}:=\sum_{i=1}^n v_i w_i^q.
\end{equation}
Again, classical codes which are self-orthogonal with respect to
(\ref{eq:Hermitian}) give rise to QECCs for $q$-dimensional systems.
\begin{corollary}\label{coro:linearQECC}\label{theorem:linearQECC}
Let $C$ be a $GF(q^2)$-linear $[n,k]_{q^2}$ self-orthogonal code over
$GF(q^2)$ and let $d=\min\{\wgt(\bm{v})\colon\bm{v}\in C^*\setminus
C\}$. Then there exists a QECC ${\cal C}=\QECC(n,n-2k,d,q)$.
\end{corollary}
\begin{proof}
From the self-orthogonal code $C$ over $GF(q^2)$ one obtains a
self-orthogonal code $D$ over $GF(q)\times GF(q)$ as follows. Let
$\gamma\in GF(q^2)\setminus GF(q)$ so that $\gamma^q=-\gamma+\gamma_0$
for some $\gamma_0\in GF(q)$. Expanding each symbol of $GF(q^2)$ with
respect to the basis $\{1,\gamma\}$ of $GF(q^2)/GF(q)$, we can write
any element $\bm{c}\in C$ as $\bm{v}+\gamma\bm{w}$ where
$\bm{v},\bm{w}\in GF(q)^n$. Then the code $D$ is defined as
$$
D:=\{(\bm{v},\bm{w})\colon \bm{v},\bm{w}\in GF(q)^n\mid\bm{v}+\gamma\bm{w}\in C\}.
$$
As $C$ is self-orthogonal with respect to (\ref{eq:Hermitian}), we get
\begin{eqnarray}
0  = \bm{c}*\bm{w}
&=&\sum_{i=1}^n(v_i+\gamma w_i)(v_i'+\gamma w_i')^q\nonumber\\
&=&\sum_{i=1}^n v_i v_i'+\gamma v_i'w_i+\gamma^q v_i w_i'+\gamma^{q+1} w_i w_i'\nonumber\\
&=&\sum_{i=1}^nv_i v_i'+\gamma^{q+1} w_i w_i'+\gamma_0 v_i w_i'+\gamma(v_i'w_i-v_i w_i').\label{eq:ip}
\end{eqnarray}
As $\gamma^{q+1}$ is the norm of $\gamma$ and hence $\gamma^{q+1}\in
GF(q)$, the coefficient $(v_i'w_i-v_i w_i')$ of $\gamma$ in
(\ref{eq:ip}) vanishes. This implies that $D$ is self-orthogonal with
respect to (\ref{eq:symplectic}). The result follows using
Theorem~\ref{theorem:additiveQECC} (see also Corollary 1 in
Ref.~\refcite{AsKn01}).
\end{proof}
Finally, the construction of so-called CSS
codes\cite{CaSh96,Ste96:error} uses the notion of duality with respect
to the Euclidean inner product
\begin{equation}\label{eq:Euclidean}
\bm{v}\cdot\bm{w}:=\sum_{i=1}^n v_i w_i,
\end{equation}
for which the dual code is denoted by $C^\bot$.
\begin{theorem}{\bfseries(CSS codes)}\label{theorem:CSS}
Let $C_1=[n,k_1,d_1]_q$ and $C_2=[n,k_2,d_2]_q$ be linear codes over
$GF(q)$ with $C_2^\bot\subseteq C_1$. Furthermore, let
$d=\min\{\wgt(\bm{v})\colon\bm{v}\in(C_1\setminus C_2^\bot)\cup
(C_2\setminus C_1^\bot)\}\ge\min(d_1,d_2)$. Then there exists a QECC
${\cal C}=\QECC(n,k_1+k_2-n,d,q)$.
\end{theorem}
\begin{proof}
It is easy to show that the code $C_1^\bot\times C_2^\bot$ is a
self-orthogonal code over $GF(q)\times GF(q)$. Applying
Theorem~\ref{theorem:additiveQECC} to this code completes the proof.
\end{proof}
In particular, Theorem~\ref{theorem:CSS} applies to so-called weakly
self-dual codes with $C\subseteq C^\bot$.
\begin{corollary}\label{theorem:WSD_CSS}
Let $C$ be an $[n,k]_q$ weakly self-dual code over $GF(q)$ and let
$d=\min\{\wgt(\bm{v})\colon\bm{v}\in C^\bot\setminus C\}$. Then there
exists a QECC ${\cal C}=\QECC(n,n-2k,d,q)$.
\end{corollary}
\begin{proof}
The results follows setting $C_1^\bot=C_2^\bot=C$ in
Theorem~\ref{theorem:CSS}. Alternatively, one can apply
Corollary~\ref{coro:linearQECC} to the self-orthogonal code $C\otimes
GF(q^2)$.
\end{proof}

Before presenting the families of classical error-correcting codes
used in our construction, we quote the quantum version of the
singleton bound:\cite{Rai99:nonbinary}
\begin{theorem}{\bfseries(Quantum Singleton Bound)} Let ${\cal
C}=\QECC(n,k,d,q)$ be a quantum error-correction code. Then
\begin{equation}\label{eq:singleton}
k+2d\le n+2.
\end{equation}
If equality holds in (\ref{eq:singleton}) then ${\cal C}$ is pure.
\end{theorem}

\begin{definition}{\bfseries(Quantum MDS code)}
A quantum code for which equality holds in (\ref{eq:singleton}), i.\,e.,
${\cal C}=\QECC(n,n-2d+2,d,q)$, is called a quantum MDS code.
\end{definition}

\section{Self-orthogonal Classical MDS Codes}
Our construction of quantum MDS codes is based on classical MDS
codes. Let $q$ be any prime power and let $\mu$, $0\le\mu<q-2$, be an
integer. By $C^{(q,\mu)}$ we denote the code generated by
\begin{equation}\label{eq:genmat}
G^{(q,\mu)}:=
\left(
\begin{array}{cccccc}
1&1&1&\ldots&1&1\\
\alpha^0&\alpha^1&\alpha^2&\ldots&\alpha^{q-2}&0\\
\alpha^0&\alpha^2&\alpha^4&\ldots&\alpha^{2(q-2)}&0\\
\vdots&\vdots&\vdots&\ddots&\vdots&\vdots\\
\alpha^0&\alpha^{\mu}&\alpha^{2\mu}&\ldots&\alpha^{\mu(q-2)}&0
\end{array}
\right),
\end{equation}
where $\alpha$ is a primitive element of $GF(q)$ and hence a primitive
$(q-1)$-th root of unity. The code $C^{(q,\mu)}$ is the dual of an
extended Reed-Solomon code. It is a maximum distance separable (MDS)
code with parameters $C^{(q,\mu)}=[q,\mu+1,q-\mu]_q$ (see
Ref.~\refcite{MS77}). Furthermore, by $C_s^{(q,\mu)}$ we denote the
code that is obtained by shortening the code $C^{(q,\mu)}$ at the last
coordinate. Again, $C_s^{(q,\mu)}=[q-1,\mu,q-\mu]_q$ is an MDS code. We
now show that both codes are contained in their duals.
\begin{lemma}\label{lemma:WSD}
For $0\le \mu <(q-1)/2$ the codes $C^{(q,\mu)}$ and $C_s^{(q,\mu)}$
are weakly self-dual with respect to the Euclidean inner product over
$GF(q)$.
\end{lemma}
\begin{proof}
It is sufficient to show that $C^{(q,\mu)}$ is contained in its dual,
i.\,e., $G^{(q,\mu)}\cdot \left(G^{(q,\mu)}\right)^t=0$.  For
$i=0,\ldots,\mu$, let $G_i$ denote the $(i+1)$-th row of
$G^{(q,\mu)}$. We have to show that the inner product $G_i\cdot G_j$
vanishes for $0\le i,j\le \mu$. Obviously, $G_0\cdot G_0=0$. If not
both $i$ and $j$ are zero, we get
\begin{equation}
G_i\cdot G_j
=\sum_{l=0}^{q-2} \alpha^{il}\alpha^{jl}
=\sum_{l=0}^{q-2} \left(\alpha^{(i+j)}\right)^l.
\end{equation}
If $i+j\not\equiv 0\bmod (q-1)$, then
\begin{equation}
G_i\cdot G_j
=\frac{\left(\alpha^{(i+j)}\right)^{q-1}-1}{\alpha^{(i+j)}-1}=0.
\end{equation}
\medskip
\end{proof}
These codes are not only weakly-self dual with respect to the
Euclidean inner product, what is more, for suitably chosen parameters
they are self-orthogonal with respect to the Hermitian inner product
as well.  This is the content of the following lemma which will
ultimately allow to define MDS codes of length $q^2$ for quantum
systems of dimension $q$.
\begin{lemma}\label{lemma:selfortho}
For $0\le \mu\le q-2$ the codes $C^{(q^2,\mu)}$ and $C_s^{(q^2,\mu)}$
are self-orthogonal with respect to the Hermitian inner product over
$GF(q^2)$.
\end{lemma}
\begin{proof}
We use the notation of the proof of Lemma~\ref{lemma:WSD}. Again
$G_0*G_0=0$. If not both $i$ and $j$ are zero, we get
\begin{equation}
G_i* G_j
=\sum_{l=0}^{q^2-2} \alpha^{il}\left(\alpha^{jl}\right)^q
=\sum_{l=0}^{q^2-2} \left(\alpha^{(i+qj)}\right)^l.
\end{equation}
So $G_i* G_j=0$ if $i+qj\not\equiv 0\bmod (q^2-1)$. This is true since $0\le
i,j\le q-2$.
\end{proof}

\section{Quantum MDS Codes}
From the classical MDS codes of the previous section one can directly
obtain quantum MDS codes.
\begin{theorem}\label{theorem:QMDS_CSS}
Let $q$ be an arbitrary prime power. Then for $0\le \mu<(q-1)/2$ there
exist quantum MDS codes with parameters
\begin{eqnarray*}
{\cal C}^{(q,\mu)}&=&\QECC(q,q-2\mu-2,\mu+2,q)\\
\mbox{and}\qquad
{\cal C}_s^{(q,\mu)}&=&\QECC(q-1,q-2\mu-1,\mu+1,q).
\end{eqnarray*}
\end{theorem}
\begin{proof}
By Lemma~\ref{lemma:WSD}, we obtain 
$C^{(q,\mu)}=[q,\mu+1,q-\mu]_q \le{C^{(q,\mu)}}^\bot$
and
$C_s^{(q,\mu)}=[q-1,\mu,q-\mu]_q\le{C_s^{(q,\mu)}}^\bot$.
As the dual of an MDS code is again an MDS code (see Theorem~2 in
Ch.~11 of Ref.~\refcite{MS77}), 
${C^{(q,\mu)}}^\bot=[q,q-\mu-1,\mu+2]_q$
and
${C_s^{(q,\mu)}}^\bot=[q-1,q-\mu-1,\mu+1]_q$.
Using the construction of Cor.~\ref{theorem:WSD_CSS}, we obtain the
quantum codes with the desired parameters.
\end{proof}
While the length of these codes is upper bounded by the dimension $q$
of the subsystems, there are also codes of length $q^2$.
\begin{theorem}\label{theorem:QMDS_Hermitean}
For any prime power $q$ and any integer $\mu$, $0\le \mu<q-1$, there
exist quantum MDS codes with parameters
\begin{eqnarray*}
{\cal D}^{(q^2,\mu)}&=&\QECC(q^2,q^2-2\mu-2,\mu+2,q)\\
\mbox{and}\qquad
{\cal D}_s^{(q^2,\mu)}&=&\QECC(q^2-1,q^2-2\mu-1,\mu+1,q).
\end{eqnarray*}
\end{theorem}
\begin{proof}
By Lemma~\ref{lemma:selfortho}, we obtain
$C^{(q^2,\mu)}=[q^2,\mu+1,q^2-\mu]_{q^2}\le{C^{(q^2,\mu)}}^*$
and
$C_s^{(q^2,\mu)}=[q^2-1,\mu,q^2-\mu]_{q^2}\le{C_s^{(q^2,\mu)}}^*$.
The dual codes have parameters
${C^{(q^2,\mu)}}^*=[q^2,q^2-\mu-1,\mu+2]_{q^2}$
and
${C_s^{(q,\mu)}}^*=[q^2-1,q^2-\mu-1,\mu+1]_{q^2}$.
We now use the construction of Cor.~\ref{theorem:linearQECC} to 
obtain quantum codes with the desired parameters.
\end{proof}

\section{Shortening Quantum Codes}\label{sec:RainsPuncturing}
While classical linear codes can be shortened to any length, i.\,e.,
from a code $[n,k,d]$ one obtains a code $[n-r,k'\ge k-r,d'\ge d]$ for
any $r$, $0\le r\le k$, this is in general not true for quantum
codes. However, in Ref.~\refcite{Rai99:nonbinary} it is shown how
quantum codes can be shortened. Here we recall the main results.
First, consider the vector valued bilinear form on $GF(q)^n\times
GF(q)^n$ defined by
\begin{equation}\label{eq:vec_form}
\{(\bm{v},\bm{w}),(\bm{v}',\bm{w}')\}:=(v_i w_i'-v_i'w_i)_{i=1}^n\in GF(q)^n.
\end{equation}
Then, for a $GF(q)$-linear code $C$ over $GF(q)\times GF(q)$, the
puncture code of $C$ is defined as
\begin{equation}\label{def:PC}
P(C):=\Bigl\langle \{\bm{c},\bm{c}'\}\colon \bm{c},\bm{c}'\in
C\Bigr\rangle^\bot\subseteq GF(q)^n,
\end{equation}
where the angle brackets denote the $GF(q)$ linear span.  From
Theorem~3 of Ref.~\refcite{Rai99:nonbinary} we get a characterization
of the shortened quantum codes which can be obtained from $C$:

\begin{theorem}\label{theorem:puncture}
Let $C$ be a subspace of $(GF(q)\times GF(q))^n$, not necessarily
self-orthogonal, of length $n$ and size $q^{n-k}$ such that $C^*$ has
minimum distance $d$. If there exists a codeword in $P(C)$ of weight
$r$, then there exists a QECC $\QECC(r,k',d',q)$ for some $k'\ge
k-(n-r)$ and $d'\ge d$.
\end{theorem}
\begin{proof}
Let $\bm{x}\in P(C)$ be a codeword of weight $r$. We define the code
$\widetilde{C}$ to be
\begin{equation}\label{eq:C_tilde}
\widetilde{C}:=\{ (\bm{v},(x_i w_i)_{i=1}^n)\colon (\bm{v},\bm{w})\in C\},
\end{equation}
i.\,e., we multiply the coordinates of the second component $\bm{w}$ by
the corresponding elements of $\bm{x}$. For arbitrary
$(\widetilde{\bm{v}},\widetilde{\bm{w}}),(\widetilde{\bm{v}}',\widetilde{\bm{w}}')\in\widetilde{C}$,
we get
\begin{equation}
\begin{array}[b]{rclcl}
(\widetilde{\bm{v}},\widetilde{\bm{w}})*(\widetilde{\bm{v}}',\widetilde{\bm{w}}')
&=&\ds\sum_{i=1}^n  \widetilde{v}_i \widetilde{w}_i'-\widetilde{v}_i'\widetilde{w}_i
&=&\ds\sum_{i=1}^n  v_i w_i' x_i-v_i'w_i x_i\\[1.5em]
&=&\ds\sum_{i=1}^n (v_i w_i'-v_i'w_i) x_i
&=&\{(\bm{v},\bm{w}),(\bm{v}',\bm{w}')\}\cdot \bm{x}.
\end{array}
\label{eq:proof_PC}
\end{equation}
From (\ref{def:PC}) it follows that (\ref{eq:proof_PC}) vanishes, i.\,e.,
$\widetilde{C}$ is self-orthogonal. As (\ref{eq:proof_PC}) depends
only on the coordinates of $\bm{x}$ that are non-zero, we can delete
the other positions in $\widetilde{C}$ and obtain a self-orthogonal
code $D\subseteq GF(q)^r\times GF(q)^r$ given by
$$
D:=\{ ((v_i),(x_i w_i))_{i\in S}\colon (\bm{v},\bm{w})\in C\},
$$
where the set $S=\{i\colon i\in \{1,\ldots,n\}\mid x_i\ne 0\}$ is
the support of the codeword $\bm{x}$. Deleting some positions, i.\,e.,
puncturing the code $\widetilde{C}$ may reduce its dimension, so $D$
has $q^{n-k'}$ codewords, for some $k'\ge k$. The dual code $D^*$ is
obtained by shortening the code $\widetilde{C}^*$. So the minimum
distance $d'$ of $D^*$ is not smaller than the minimum distance of
$\widetilde{C}^*$ which is at least as large as that of $C^*$. This
shows $d'\ge d$.
\end{proof}

In order to apply Theorem~\ref{theorem:puncture} to our codes, we
study the puncture code. For the codes of CSS type, we have the
following:
\begin{theorem}\label{theorem:PC_CSS}
Let $C=C_1^\bot\times C_2^\bot\subseteq GF(q)^n\times GF(q)^n$ as in
Theorem~\ref{theorem:CSS}. Then
\begin{equation}
P(C)=\Bigl\langle (c_i d_i)_{i=1}^n\colon \bm{c}\in C_1^\bot, \bm{d}\in C_2^\bot\Bigl\rangle^\bot.
\end{equation}
\end{theorem}
\begin{proof}
In order to find generators of $P(C)$, it suffices to compute the
bilinear form (\ref{eq:vec_form}) for all pairs of elements of a
vector space basis for $C$. Using the basis $ \{(\bm{c},\bm{0})\colon
\bm{c}\in C_1^\bot\} \cup \{(\bm{0},\bm{d})\colon \bm{d}\in C_2^\bot\}
$, the result follows.
\end{proof}
For a $GF(q^2)$-linear code $C$ over $GF(q^2)$, the situation is a bit
more complicated. The following theorem shows how to compute $P(C)$ in
this case:
\begin{theorem}\label{theorem:PC_linear}
Let $C$ be a $GF(q^2)$-linear code. Then
\begin{equation}\label{def:PC_linear}
P(C)=\Bigl\langle (c_i d_i^q+c_i^q d_i)_{i=1}^n\colon \bm{c},\bm{d}\in C\Bigl\rangle^\bot.
\end{equation}
\end{theorem}
\begin{proof}
Similar to the proof of Theorem~\ref{theorem:puncture}, we will show
that each codeword of $P(C)$ defined by (\ref{def:PC_linear}) gives
rise to a shortened quantum code. First note that
$$
c_i d_i^q+c_i^q d_i=
c_i d_i^q+(c_i d_i^q)^q=\tr(c_i d_i^q),
$$
where $\tr\colon GF(q^2)\rightarrow GF(q), x\mapsto x+x^q$ denotes the
trace of the field extension $GF(q^2)/GF(q)$. Hence $P(C)$ is the dual
code of the component-wise trace of the code generated by
$\bigl\langle (c_i d_i^q)_{i=1}^n\colon \bm{c},\bm{d}\in
C\bigr\rangle$.  As the dual of the trace code equals the restriction
of the dual code to the subfield, i.\,e.,
$(\tr_K(C))^\bot=(C^\bot)|_K$ (see, e.\,g., Theorem~11 in Ch.~7, \S~7
of Ref.~\refcite{MS77}), we can rewrite (\ref{def:PC_linear}) as
\begin{equation}\label{eq:PC_linear2}
P(C)=\Bigl\langle (c_i d_i^q)_{i=1}^n\colon
\bm{c},\bm{d}\in C\Bigl\rangle^\bot \mathrel{\cap} GF(q)^n.
\end{equation}
As in the proof of Cor.~\ref{coro:linearQECC}, we expand each codeword
$\bm{c}\in C$ as $\bm{c}=\bm{v}+\gamma\bm{w}$ where $\gamma\in
GF(q^2)\setminus GF(q)$ with $\gamma^q=-\gamma+\gamma_0$ for some
$\gamma_0\in GF(q)$.  This defines a $GF(q)$-linear code over
$GF(q)\times GF(q)$ which is given by
$$
D=\{(\bm{v},\bm{w})\colon \bm{v},\bm{w}\in GF(q)^n\mid\bm{v}+\gamma\bm{w}\in C\}.
$$
Then, as in  (\ref{eq:C_tilde}), for a codeword $\bm{x}\in P(C)$ of weight
$r$, we define the code
$$
\widetilde{D}:=\{ (\bm{v},(x_i w_i)_{i=1}^n)\colon \bm{v},\bm{w}\in GF(q)^n\mid\bm{v}+\gamma\bm{w}\in C\}.
$$
From (\ref{eq:PC_linear2}) it follows that $\sum_{i=1}^n x_i c_i
d_i^q$ vanishes. Similar to the proof of Cor.~\ref{coro:linearQECC}
(see eq.~(\ref{eq:ip})), this implies that $\widetilde{D}$ is
self-orthogonal with respect to (\ref{eq:symplectic}), as well as the
code obtained by deleting all coordinates where $\bm{x}$ is zero.
\end{proof}

\section{Results}
Applying Theorem~\ref{theorem:PC_CSS} and
Theorem~\ref{theorem:PC_linear} to the codes of Lemma~\ref{lemma:WSD}
and Lemma~\ref{lemma:selfortho}, respectively, we obtain
\begin{eqnarray}
P(C^{(q,\mu)})&=&
\Bigl\langle G_{i+j}\colon 0\le i,j\le\mu\Bigr\rangle^\bot=
{C^{(q,2\mu)}}^\bot\label{eq:PC_CSS}\\
\mbox{and}\qquad
P(C^{(q^2,\mu)})&=&\Bigl\langle G_{i+q j}\colon 0\le
i,j\le\mu\Bigr\rangle^\bot,\label{eq:PC_selfortho}
\end{eqnarray}
where again $G_i$ denotes the $(i+1)$-th row of the matrix
$G^{(q,\mu)}$ in (\ref{eq:genmat}). Additionally, we have used that
the component-wise product of $G_i$ and $G_j$ is $G_{i+j}$.

In combination with Theorem~\ref{theorem:puncture}, we finally get:
\begin{theorem}
Let $q$ be an arbitrary prime power. Then for all $3\le n\le q$ and
$1\le d\le n/2+1$ there exists quantum MDS codes
$\QECC(n,n-2d+2,d,q)$.  Moreover, for $2\le d\le q$ and some $s$ (at
least $s=0$ and $s=1$) there exist quantum MDS codes
$\QECC(q^2-s,q^2-2d+2-s,d,q)$.
\end{theorem}
\begin{proof}
The puncture code (\ref{eq:PC_CSS}) is again an MDS code. As MDS codes
contain words of all weights $d_{\text{min}}\le w\le n$ from the
minimum distance $d_{\text{min}}$ to the length $n$ of the
code\cite{MS77}, shortening of the corresponding quantum MDS code to
any length with the obvious constraints is possible.

For the codes of length $q^2$ from Lemma~\ref{lemma:selfortho}, we do
not have an explicit formula for the weights in $P(C)$, but from
Theorem~\ref{theorem:QMDS_Hermitean} one knows that at least quantum
MDS codes of length $q^2$ and $q^2-1$ exist.
\end{proof}

The preceding theorem does not give much information about quantum
MDS codes of length $n$ with $q<n<q^2-1$.  For a specific code,
however, one can compute the puncture code $P(C)$ using
(\ref{eq:PC_selfortho}). It can also be shown that in that case $P(C)$
is an extended cyclic code, but in general not an MDS code. Hence it
is difficult to compute its weight distribution, especially for large
codes for which only random sampling is possible. Using the computer
algebra system {\sf MAGMA}\cite{Magma}, we have computed and studied
$P(C)$ for quantum MDS codes for quantum systems of dimension
$q\in\{2,3,4,5,7\}$. The results which are summarized in
Table~\ref{table:results} indicate that many shortenings are possible.

\begin{table}
\tbl{Possible shortenings of QECCs of length $q^2$ for quantum systems
of dimension $q$. Note that e.\,g. for the code $\QECC(16,10,4,4)$,
there are only words of even weight in $P(C)$, and for the code
$\QECC(25,19,4,5)$, there is no codeword of weight $7$ in
$P(C)$. Hence e.\,g. codes $\QECC(7,1,4,4)$ and $\QECC(7,1,4,5)$
cannot be obtained directly via shortening (but at least a code
$\QECC(7,1,4,5)$ can be constructed by other methods).
\label{table:results}}{
\begin{minipage}{\textwidth}
\let\itshape\relax
\def\arraystretch{1.5}
\arraycolsep2\arraycolsep
$$\begin{array}{c|c|c|l}
q&
\mbox{Theorem~\ref{theorem:QMDS_Hermitean}} &
\mbox{puncture code $P(C)$} & \mbox{weights in $P(C)$}\\
\hline
2&\QECC(4,2,2,2)&[4,3,2]_2&2,4\\
\hline
3&\QECC(9,7,2,3)&[9,8,2]_3&2\text{--}9\\
 &\QECC(9,5,3,3)&[9,5,4]_3&4\text{--}9\\
\hline
4&\QECC(16,14,2,4)&[16,15,2]_4&2\text{--}16\\
 &\QECC(16,12,3,4)&[16,12,4]_4&4\text{--}16\\
 &\QECC(16,10,4,4)&[16,7,8]_4&8,10,12,14,16\\
\hline
5&\QECC(25,23,2,5)&[25,24,2]_5&2\text{--}25\\
 &\QECC(25,21,3,5)&[25,21,4]_5&4\text{--}25\\
 &\QECC(25,19,4,5)&[25,16,6]_5&6,8\text{--}25\\
 &\QECC(25,17,5,5)&[25,9,12]_5&12\text{--}25\\
\hline
7&\QECC(49,47,2,7)&[49,48,2]_7 &2\text{--}49\\
 &\QECC(49,45,3,7)&[49,45,4]_7 &4\text{--}49\\
 &\QECC(49,43,4,7)&[49,40,6]_7 &6\text{--}49\\
 &\QECC(49,41,5,7)&[49,33,8]_7 &8,12\text{--}49\\
 &\QECC(49,39,6,7)&[49,24,16]_7&16,18\text{--}49\mbox{\footnote{A codeword
of weight 17 might exists as well, but the code $[49,24,16]_7$ has too
many codewords for complete enumeration.}}\\
 &\QECC(49,37,7,7)&[49,13,24]_7&24,25,28,30\text{--}49
\magma{WeightDistribution
[ <0, 1>, <24, 7644>, <25, 7056>, <28, 142296>, <30, 1575840>, <31,
5496624>, <32, 23020200>, <33,  59136336>, <34, 165251520>, <35,
440823600>, <36, 1023564528>, <37, 2129618400>, <38, 4135698000>, <39,
6839754768>, <40, 10446702000>, <41, 13517493192>, <42, 15657866448>,
<43, 15237911808>, <44, 12445916112>, <45, 8289582840>, <46,
4351416384>, <47, 1647145584>, <48, 421778868>, <49, 49100358> ]
}
\end{array}
$$
\end{minipage}}
\end{table}

\section{Final Remarks}
Following the presentation of these results at the conference
EQIS\,'03, we have learned about the work of Chi et al.\cite{CKLL01}
The authors constructed also quantum MDS codes, but only for quantum
systems of odd dimension $p^m$, where $p$ is a prime, and maximal
length $p^m$. Our constructions apply to both even and odd prime
powers. Moreover, we obtain quantum MDS codes for quantum systems of
dimension $p^m$ of length up to $p^{2m}$.

Finally, we note that we have found a generalization of our
constructions that increases the maximal length of the resulting codes
by one, i.\,e., up to $p^{2m}+1$.  It remains an open question what
the maximal length $n$ of a non-trivial quantum MDS code with minimum
weight $d>2$ is.

\section*{Acknowledgments}
Markus Grassl and Martin R\"otteler acknowledge the hospitality of the
Mathematical Sciences Research Institute, Berkeley, USA, where part of
this work has been performed. M.~R. was supported in part by CFI,
ORDFC, and MITACS.  Funding by {\em Deutsche Forschungsgemeinschaft
(DFG), Schwerpunktprogramm Quanten-Informationsverarbeitung (SPP
1078), Projekt AQUA (Be~887/13)} is also acknowledged.

\end{document}